# Interacting dark matter and *q*-deformed dark energy non-minimally coupled to gravity

## Emre Dil

Department of Physics, Sinop University, Korucuk, 57000-Sinop, Turkey
e-mail: emredil@sakarya.edu.tr

**Abstract:** In this paper, we propose a new approach to study the dark sector of the universe by considering the dark energy as emerging a *q*-deformed bosonic scalar field which is not only interacting with the dark matter, but also non-minimally coupled to gravity, in the framework of standard Einsteinian gravity. In order to analyze the dynamic of the system, we first give the quantum field theoretical description of the *q*-deformed scalar field dark energy, then construct the action and the dynamical structure of these interacting and non-minimally coupled dark sector. As a second issue, we perform the phase space analysis of the model to check the reliability of our proposal by searching the stable attractor solutions implying the late-time accelerating expansion phase of the universe.

## 1. Introduction

The dark energy is accepted as the effect of causing the late-time accelerated expansion of universe which is experienced by the astrophysical observations such that, Supernova Ia [1,2], large scale structure [3,4], the baryon acoustic oscillations [5] and cosmic microwave background radiation [6-9]. According to the standard model of cosmological data 70% of the content of the universe consists of dark energy. Moreover, the remaining 25% of the content is an unknown form of matter having a mass but in non-baryonic form that is called as dark matter and the other 5% of the energy content of the universe belongs to ordinary baryonic matter [10]. While the dark energy spread all over the intergalactic media of the universe and produces a gravitational repulsion by its negative pressure to drive the accelerating expansion of the universe, the dark matter is distributed over the inner galactic media inhomogeneously and it contributes to the total gravitational attraction of the galactic structure and fixes the estimated motion of galaxies and galactic rotation curves [11, 12].

Miscellaneous dark models have been proposed to explain a better mechanism for the accelerated expansion of the universe. These models include interactions between dark energy, dark matter and the gravitational field. The coupling between dark energy and dark matter seems possible due to the equivalence of order of the magnitudes in the present time [13-22]. On the other hand, there are also models in which the dark energy non-minimally couples to gravity in order to provide quantum corrections and renormalizability of the scalar field in the curved spacetime. Also the crossing of the dark energy from the quintessence phase to phantom phase, known as the Quintom scenario, can be possible in the models that the dark energy interacts with the gravity. If the dark energy minimally couples to gravity, the equation of state parameter of the dark energy cannot cross the cosmological constant boundary $\omega =1$ in the Friedmann-Robertson-Walker (FRW) geometry; therefore it is possible to emerge the Quintom scenario in the model that the dark energy non-minimally couples to gravity [23-37].

The constitution of the dark energy can be alternatively the cosmological constant $\Lambda$ with a constant energy density filling the space homogeneously [38-41]. As the varying energy density dark energy models, instead of the cosmological constant, quintessence, phantom and tachyon fields can be considered. However, all these different dark energy models are same in terms of the non-deformed field constituting the dark energy. There is no reason to prevent us assuming the dark energy as a deformed scalar field, having a negative pressure, too, as required the dark energy has to be. Therefore, we propose that the dark energy considered in this study is formed of the deformed scalar field whose field equations are defined by the deformed oscillator algebras.

The quantum algebra and quantum group structure were firstly introduced by Kulish, Reshetikhin, Sklyanin, Takhtajan and Faddeev [42-45] during the investigations of integrable systems in quantum field theory and statistical mechanics. Quantum groups and deformed boson algebras are closely related terms. It is known that the deformation of the standard boson algebra is first proposed by Arik-Coon [46]. Later on, Macfarlane and Biedenharn have realized the deformation of boson algebra in a different manner from Arik-Coon [47, 48]. The relation between quantum groups and the deformed oscillator algebras can be constructed obviously with this study by expressing the deformed boson operators in terms of the $su_q(2)$ Lie algebra operators. Therefore, the construction of the relation between quantum groups and deformed algebras leads the deformed algebras of great interest with many different applications. The deformed version of Bardeen-Cooper-Schrieffer (BCS) many-body formalism in nuclear force, deformed creation and annihilation operators are used to study the quantum occupation probabilities [49]. As another study, in Nambu-Jona-Lasinio (NJL) model the deformed fermion operators are used instead of standard fermion operators and this leads an increase in the NJL four-fermion coupling force and the quark condensation related to the dynamical mass [50]. The statistical mechanical studies of the deformed boson and fermion systems have been familiar in recent years [51-61]. Moreover, the investigations on the internal structure of composite particles involve the deformed fermions or bosons as the building block of the composite structures [62, 63]. There are also applications of the deformed particles in black hole physics [64-67]. The range of the deformed boson and fermion applications diverse from atomic-molecular physics to solid state physics in a widespread manner [68-73].

The ideas on considering the dark energy as the deformed scalar field have become common in the literature [74-77]. In this study, we then take into account the deformed bosons as the scalar field dark energy interacting with the dark matter and also non-minimally coupled to gravity. In order to confirm our proposal that the dark energy can be considered as a deformed scalar field, we firstly introduce the dynamics of the interacting and non-minimally coupled dark energy, dark matter and gravity model in a spatially-flat FRW background, then perform the phase space analysis to check whether it will provide the late-time stable attractor solutions implying the accelerated expansion phase of the universe.

## 2. Dynamics of the model

The field equations of the scalar field dark energy are considered to be defined by the $q$-deformed boson fields in our model. Constructing a $q$-deformed quantum field theory after the idea of $q$-deformation of the single particle quantum mechanics [46-48] has naturally been non-surprising [78-80]. The bosonic part of the deformed particle fields corresponds to the deformed scalar field and the fermionic counterpart corresponds to the deformed vector field. In this study, we consider the $q$-deformed bosonic scalar field as the $q$-deformed dark energy under consideration. In our model, the $q$-deformed dark energy interacts with the dark matter and also non-minimally couples to gravity.

Early Universe scenarios can be well understood by studying the quantum field theory in curved space-time. The behavior of the classical scalar field near the initial singularity can be translated to the quantum field regime by constructing the coherent states in quantum mechanics for any mode of the scalar field. It is now impossible to determine the quantum state of the scalar field near the initial singularity by an observer, at the present universe. In order to overcome the undeterministic nature, Hawking proposes to take the random superposition of all possible states in that space-time. It has been realized by Berger with taking random superposition of coherent states. Also the particle creation in an expanding universe with a non-quantized gravitational metric has been investigated by Parker. It has been stated by Goodison and Tom that if the field quanta obey the Bose or Fermi statistics, when considered the evolution of the scalar field in an expanding universe, then the particle creation does not occur in the vacuum state. Their result gives signification the possibility of the existence of the deformed statistics in coherent or squeezed states in the Early Universe [80-85].

Motivated by this significant possibility, we propose that the dark energy consists of a $q$-deformed scalar field whose particles obey the $q$-deformed algebras. Therefore, we now define the $q$-deformed scalar field constructing the dark energy in our model. The field operator of the $q$-deformed scalar field dark energy can be given as [80]

$$\phi_q(x) = \int \frac{d^3k}{(2\pi)^{3/2}} \frac{1}{(2w_k)^{1/2}} \left[ a_q(k)e^{ikx} + a_q^*(k)e^{-ikx} \right], \tag{1}$$

The following commutation relations for the deformed annihilation operator $a_q(k)$ and creations operator $a_q^*(k)$ in $q$-bosonic Fock space are given by [46]

$$a_q(k)a_q^*(k') - q^2 a_q^*(k')a_q(k) = \delta(k-k'), \tag{2}$$

$$a_q(k)a_q(k') - q^2 a_q(k')a_q(k) = 0, \tag{3}$$

where $q$ is a real deformation parameter in interval $0 < q < \infty$ and $[\hat{N}(k)] = a_q^*(k)a_q(k)$ is the deformed number operator of $k$-th mode whose eigenvalue spectrum is given as

$$[N(k)] = \frac{1-q^{2N(k)}}{1-q^2}. \tag{4}$$

Here $\hat{N}(k) = a_s^*(k)a_s(k)$ is the standard non-deformed number operator. By using (2) and (3) in (1), we can obtain the commutation relations and planewave expansion of the $q$-deformed scalar field $\phi_q(x)$, as follows

$$\phi_q(x)\phi_q^*(x') - q^2\phi_q^*(x')\phi_q(x) = i\Delta(x-x'), \tag{5}$$

where

$$\Delta(x-x') = \frac{-1}{(2\pi)^3} \int \frac{d^3k}{w_k} \sin w_k(x-x_0). \tag{6}$$

The metric of the spatially flat FRW space-time in which the $q$-oscillator algebra represents the $q$-deformed scalar field dark energy is defined by

$$ds^2 = -dt^2 + a^2(t)[dr^2 + r^2 d\theta^2 + r^2 \sin^2\theta \, d\phi^2], \tag{7}$$

and for a FRW metric

$$w_k^2 = g\left(\sum_i \frac{k_i^2}{a^2} + m\right), \tag{8}$$

where $g = \det g_{\mu\nu}$. Also the relation between deformed and standard annihilation operators $a_q$ and $a_s$ [86] is given as

$$a_q = a_s \sqrt{\frac{[\hat{N}]}{\hat{N}}}, \tag{9}$$

which is used to obtain the relation between deformed and standard bosonic scalar fields by using (4) in (9) and (1):

$$\phi_q = \phi \sqrt{\frac{1-q^{2\hat{N}}}{(1-q^2)\hat{N}}}. \tag{10}$$

Here we have used the Hermiticity of the number operator $\hat{N}$.

Now the Friedmann equations will be derived for our interacting dark matter and non-minimally coupled $q$-deformed dark energy model in a FRW space-time by using the scale factor $a(t)$ in Einstein's equations. In order to obtain these equations, we relate the scale factor to the energy-momentum tensor of the objects in the model under consideration. We use the fluid description of the objects in our model by considering energy and matter as a perfect fluid, which are dark energy and matter in our model. An isotropic fluid in one coordinate frame leads to an isotropic metric in another frame coinciding with the frame of the fluid. This means that the fluid is at rest in commoving coordinates. Then the four velocity of the fluid is given as [53]

$$U^{\mu} = (1,0,0,0), \tag{11}$$

and the energy-momentum tensor follows as

$$T_{\mu\nu} = (\rho + p)U_{\mu}U_{\nu} + p\, g_{\mu\nu} = \begin{pmatrix} \rho & 0 & 0 & 0 \\ 0 & & & \\ 0 & & g_{ij}p & \\ 0 & & & \end{pmatrix}. \tag{12}$$

A more suitable form can be obtained by raising one, such that

$$T^{\mu}_{\nu} = diag(-\rho, p, p, p). \tag{13}$$

Since we have two constituents, $q$-deformed dark energy and the dark matter in our model, the total energy density and the pressure are given by

$$\rho_{tot} = \rho_q + \rho_m, \tag{14}$$

$$p_{tot} = p_q + p_m, \tag{15}$$

where $\rho_q$ and $\rho_q$ are the energy density and the pressure of the $q$-deformed dark energy and the $\rho_m$ and $\rho_m$ are the energy density and the pressure of the dark matter, respectively. The equation of state of the energy-momentum carrying cosmological fluid component under consideration in the FRW universe is given by $p = \omega \rho$ which relates the pressure and the energy density and $\omega$ is called as the equation of state parameter. We then express the total equation of state parameter, such that

$$\omega_{tot} = \frac{p_{tot}}{\rho_{tot}} = \omega_q \Omega_q + \omega_m \Omega_m. \tag{16}$$

where $\Omega_q = \rho_q / \rho_{tot}$ and $\Omega_m = \rho_m / \rho_{tot}$ are the density parameters for the $q$-deformed dark energy and the dark matter, respectively. Then total the density parameter is defined as

$$\Omega_{tot} = \Omega_q + \Omega_m = \frac{\kappa^2 \rho_{tot}}{3H^2} = 1. \tag{17}$$

We now turn to Einstein's equations of the form $R_{\mu\nu} = \kappa^2 (T_{\mu\nu} - \tfrac{1}{2} g_{\mu\nu} T)$. Then by using the components of the Ricci tensor for a FRW space-time (7) and the energy momentum tensor in (13), we rewrite the Einstein's equations, for $\mu\nu = 00$ and $\mu\nu = ij$ as follows

$$-3\frac{\ddot{a}}{a} = \frac{\kappa^2}{2}(\rho + 3p), \tag{18}$$

$$\frac{\ddot{a}}{a} + 2\left(\frac{\dot{a}}{a}\right)^2 = \frac{\kappa^2}{2}(\rho - p), \tag{19}$$

respectively. Here dot also represents the derivative with respect to cosmic time $t$. Using (18) and (19) gives the Friedmann equations for the FRW metric as

$$H^2 = \frac{\kappa^2}{3}(\rho_q + \rho_m), \tag{20}$$

$$\dot{H} = -\frac{\kappa^2}{2}(\rho_q + p_q + \rho_m + p_m), \tag{21}$$

where $H = \dot{a}/a$ is the Hubble parameter. From the conservation of energy, we can obtain the continuity equations for the $q$-deformed dark energy and the dark matter constituents in the form of evolution equations, such as

$$\dot{\rho}_q + 3H(\rho_q + p_q) = -Q, \tag{22}$$

$$\dot{\rho}_m + 3H(\rho_m + p_m) = Q, \tag{23}$$

where $Q$ is an interaction current between the $q$-deformed dark energy and the dark matter which transfers the energy and momentum from the dark matter to dark energy and vice versa. $Q$ vanishes for the models having no interaction between the dark energy and the dark matter.

Now we will define the Dirac-Born-Infeld type action integral of the interacting dark matter and $q$-deformed dark energy non-minimally coupled to gravity in the framework of Eisteinian general relativity [88-90]. After that we will obtain the energy-momentum tensor $T_{\mu\nu}$ for the $q$-deformed dark energy and the dark matter in order to get the energy density $\rho$ and pressure $p$ of these dark objects explicitly. Then the action is given as

$$S = \int d^4x \sqrt{-g} \left[ \frac{R}{2\kappa^2} - \frac{1}{2} g^{\mu\nu} \partial_\mu \phi_q \partial_\nu \phi_q - V - \xi f(\phi_q) R + L_m \right], \tag{24}$$

where $\xi$ is a dimensionless coupling constant between $q$-deformed dark energy and the gravity, so $\xi f(\phi_q) R$ denotes the explicit non-minimal coupling between energy and the gravity. Also $L_q = -(1/2) g^{\mu\nu} \partial_\mu \phi_q \partial_\nu \phi_q - V - \xi f(\phi_q) R$ and $L_m$ are the Lagrangian densities of the $q$-deformed dark energy and the dark matter, respectively. Then the energy-momentum tensors of the dark energy constituent of our model can be calculated, as follows [91]

$$\begin{aligned} T_{\mu\nu}^q &= -2 \frac{\partial L_q}{\partial g^{\mu\nu}} + g_{\mu\nu} L_q \\ &= \partial_\mu \phi_q \partial_\nu \phi_q + 2\xi f(\phi_q) \frac{\partial R}{\partial g^{\mu\nu}} - \frac{1}{2} g_{\mu\nu} [g^{\alpha\beta} \partial_\alpha \phi_q \partial_\beta \phi_q + 2V] - g_{\mu\nu} \xi f(\phi_q) R. \end{aligned} \tag{25}$$

In order to find the derivative of the Ricci scalar with respect to the metric tensor, we use the variation of the contraction of the Ricci tensor identity $\delta R = R_{\mu\nu} \delta g^{\mu\nu} + g^{\mu\nu} \delta R_{\mu\nu}$. This leads us to find the variation of the contraction of the Riemann tensor identity, as follows $\delta R_{\mu\nu} = \delta R^\rho_{\mu\rho\nu} = \nabla_\rho(\delta\Gamma^\rho_{\nu\mu}) - \nabla_\nu(\delta\Gamma^\rho_{\rho\mu})$. Here $\nabla_\mu$ represents the covariant derivative and $\Gamma^\rho_{\nu\mu}$ does the Christoffel connection. By using the metric compatibility and the tensor nature of $\delta\Gamma^\rho_{\nu\mu}$, we finally obtain

$$\frac{\delta R}{\delta g^{\mu\nu}} = R_{\mu\nu} + g_{\mu\nu}\Box - \nabla_\mu\nabla_\nu, \tag{26}$$

where $\Box = g^{\alpha\beta}\nabla_\alpha\nabla_\beta$ is the covariant d'Alembertian. Using (26) in (25) gives

$$\begin{aligned} T^q_{\mu\nu} &= \partial_\mu\phi_q\partial_\nu\phi_q - \frac{1}{2}g_{\mu\nu}[g^{\alpha\beta}\partial_\alpha\phi_q\partial_\beta\phi_q] - g_{\mu\nu}V \\ &+ 2\xi[R_{\mu\nu} - \frac{1}{2}g_{\mu\nu}R]f(\phi_q) + 2\xi\Box f(\phi_q) - 2\xi\nabla_\mu\nabla_\nu f(\phi_q). \end{aligned} \tag{27}$$

Then the $\mu\nu = 0,0$ component of the energy-momentum tensor leads to the energy density $\rho_q$

$$\rho_q = T^q_{00} = \frac{1}{2}\dot\phi_q^2 + V + 6\xi H^2 f(\phi_q) + 6\xi H f'(\phi_q)\dot\phi_q, \tag{28}$$

where prime refers to derivative with respect to the field $\phi_q$ and we use $\Box = -\partial_0^2 - 3H\partial_0$, because of the homogeneity and the isotropy for $\phi_q$ in space. Also $R_{00} = -3\ddot a/a$ and $R = 6[\ddot a/a + \dot a^2/a^2]$ is used for the FRW geometry. The $\mu\nu = i,i$ components of $T^q_{\mu\nu}$ also gives the pressure $p_q$ as

$$p_q = g^{ii}T^q_{ii} = \frac{1}{2}\dot\phi_q^2 - V - 2\xi[2\dot H f(\phi_q) + 3H^2 f(\phi_q) + f''(\phi_q)\dot\phi_q^2 + f'(\phi_q)\ddot\phi_q + 2H f'(\phi_q)\dot\phi_q], \tag{29}$$

where we use $(\nabla_i\nabla_i)f(\phi_q) = (\partial_i\nabla_i - \Gamma^\lambda_{ii}\nabla_\lambda)f(\phi_q) = -\Gamma^0_{11}\partial_0 f(\phi_q)$ with $\Gamma^0_{11} = \dot a a$ for the FRW spacetime. We can now obtain the equation of motion for the $q$-deformed dark energy by inserting (28) and (29) into the evolution equation (22), such that

$$\ddot{\phi}_q + 3H\dot{\phi}_q + \frac{\partial V}{\partial \phi_q} + \xi R f'(\phi_q) = -\frac{Q}{\dot{\phi}_q}. \tag{30}$$

The usual assumption in the literature is to consider the coupling function as $f(\phi_q) = \phi_q^2/2$ [92] and the potential as $V = V_0 e^{-\kappa\lambda\phi_q}$ [93-95]. In order to find the energy density, pressure and equation of motion in terms of the deformation parameter $q$, we use the above coupling function and potential with the rearrangement of equation (10) as $\phi_q = \Delta(q)\phi$ in the equations (28)-(30) and obtain,

$$\rho_q = \frac{1}{2}\Delta^2\dot{\phi}^2 + e^{-\kappa\lambda\Delta\phi} + 3\xi H^2\Delta^2\phi^2 + 6\xi H\Delta^2\phi\dot{\phi} + \frac{1}{2}\dot{\Delta}^2\phi^2 + \Delta\dot{\Delta}\phi\dot{\phi} + 6\xi H\Delta\dot{\Delta}\phi^2, \tag{31}$$

$$\begin{aligned} p_q &= \frac{1}{2}\Delta^2\dot{\phi}^2 - e^{-\kappa\lambda\Delta\phi} - 2\xi\Delta^2[\dot{H}\phi^2 + \frac{3}{2}H^2\phi^2 + \dot{\phi}^2 + \phi\ddot{\phi} + 2H\phi\dot{\phi}] \\ &+ (1-8\xi)\Delta\dot{\Delta}\phi\dot{\phi} + (\frac{1}{2}-2\xi)\dot{\Delta}^2\phi^2 - 2\xi\Delta\ddot{\Delta}\phi^2 - 4\xi H\phi\dot{\phi}\Delta\dot{\Delta}\phi^2 \end{aligned}, \tag{32}$$

$$\Delta\ddot{\phi} + 3\Delta H\dot{\phi} - \kappa\lambda e^{-\kappa\lambda\Delta\phi} + \xi\Delta R\phi. + 2\dot{\Delta}\dot{\phi} + \ddot{\Delta}\phi + 3H\dot{\Delta}\phi = -\beta\kappa\rho_m. \tag{33}$$

Here we consider that the particles in each mode can vary by creation or annihilation in time for $\Delta = \sqrt{(1-q^{2N})/(1-q^2)N}$, therefore its time derivatives are non-vanishing. On the other hand, the common interaction current in the literature $Q = \beta\kappa\rho_m\dot{\phi}_q$ is used here [17].

Now the phase-space analysis for our interacting dark matter and non-minimally coupled $q$-deformed dark energy model will be performed, whether the late-time stable attractor solutions can be obtained, in order to confirm our model.

## 3. Phase-space and stability analysis

The cosmological properties of the proposed $q$-deformed dark energy model can be investigated by performing the phase-space analysis. Therefore, we first transform the equations of the dynamical system into its autonomous form by introducing the auxiliary variables [15, 96-100], such as

$$x = \frac{\kappa \Delta \dot\phi}{\sqrt{6} H} = \Delta x_s ,$$

$$y = \frac{\kappa \sqrt{e^{-\kappa \lambda \Delta \phi}}}{\sqrt{3} H} = \sqrt{e^{-\kappa \lambda \phi (\Delta - 1)}} \, y_s ,$$

$$z = \frac{\kappa \dot{\Delta \phi}}{\sqrt{6} H} , \quad z_s = 0$$

$$u = \kappa \Delta \phi = \Delta u_s , \tag{34}$$

where $x_s$, $y_s$, $z_s$ and $u_s$ are the standard form of the auxiliary variables in $q \to 1$ limit. We now write the density parameters for the dark matter and $q$-deformed scalar field dark energy in the autonomous system by using (31) with (39)

$$\Omega_m = \frac{\kappa^2 \rho_m}{3 H^2} , \tag{35}$$

$$\Omega_q = \frac{\kappa^2 \rho_q}{3 H^2} = x^2 + y^2 + \xi u^2 + 2\sqrt{6} \, \xi \, xu + z^2 + 2xz + 2\sqrt{6} \, \xi \, zu . \tag{36}$$

Then the total density parameter reads, as follows

$$\Omega_{tot} = \frac{\kappa^2 \rho_{tot}}{3 H^2} = x^2 + y^2 + \xi u^2 + 2\sqrt{6} \, \xi \, xu + z^2 + 2xz + 2\sqrt{6} \, \xi \, zu + \Omega_m = 1. \tag{37}$$

We should also obtain the $\kappa^2 p_q / 3 H^2$ in the autonomous form to write the equation of state parameters, such that

$$\frac{\kappa^2 p_q}{3 H^2} = (1 - 4\xi) x^2 - y^2 + (\frac{2}{3} \xi + 4 \xi^2) su^2 + (8 \xi^2 - \xi) u^2 + \frac{2\sqrt{6}}{3} \xi \, xu + (1 - 4\xi) z^2$$
$$+ 2(1 - 4\xi) xz + \frac{2\sqrt{6}}{3} \xi \, zu + 2 \xi \beta u \Omega_m - 2 \xi \lambda \, y^2 u \tag{38}$$

where $s = -\dot{H}/H^2$. Using (36) and (38), we find the equation of state parameter for the dark energy as

$$\omega_q = \frac{p_q}{\rho_q} = \left[(1-4\xi)x^2 - y^2 + (\frac{2}{3}\xi + 4\xi^2)su^2 + (8\xi^2 - \xi)u^2 + \frac{2\sqrt{6}}{3}\xi xu + (1-4\xi)z^2 \right.$$
$$\left. + 2(1-4\xi)xz + \frac{2\sqrt{6}}{3}\xi zu + 2\xi\beta u\Omega_m - 2\xi\lambda y^2 u \right]\left[x^2 + y^2 + \xi u^2 \right.$$
$$\left. + 2\sqrt{6}\,\xi xu + z^2 + 2xz + 2\sqrt{6}\,\xi zu \right]^{-1} . \quad (39)$$

Also from (36) and (39), the total equation of state parameter can be obtained as

$$\omega_{tot} = \omega_q \Omega_q + \omega_m \Omega_m = (1-4\xi)x^2 - y^2 + (\frac{2}{3}\xi + 4\xi^2)su^2 + (8\xi^2 - \xi)u^2$$
$$+ \frac{2\sqrt{6}}{3}\xi xu + (1-4\xi)z^2 + 2(1-4\xi)xz + \frac{2\sqrt{6}}{3}\xi zu , \quad (40)$$
$$+ 2\xi\beta u\Omega_m - 2\xi\lambda y^2 u + (\gamma - 1)\Omega_m$$

where $\gamma = 1 + \omega_m$ is defined to be the barotropic index. We need to give the junk parameter $s$ in the autonomous form, such as

$$s = -\frac{\dot{H}}{H^2} = \frac{3}{2}(1 + \omega_{tot}) = \frac{3}{2}\left[1 + (1-4\xi)x^2 - y^2 + (\frac{2}{3}\xi + 4\xi^2)su^2 + (8\xi^2 - \xi)u^2 \right.$$
$$\left. + \frac{2\sqrt{6}}{3}\xi xu + (1-4\xi)z^2 + 2(1-4\xi)xz + \frac{2\sqrt{6}}{3}\xi zu \right. \quad (41)$$
$$\left. + 2\xi\beta u\Omega_m - 2\xi\lambda y^2 u + (\gamma - 1)\Omega_m \right]$$

Pulling $s$ from right hand side of (41) to the left hand side gives

$$s = \left[ 1 + (1-4\xi)x^2 - y^2 + (8\xi^2 - \xi)u^2 + \frac{2\sqrt{6}}{3}\xi xu \right.$$
$$+ (1-4\xi)z^2 + 2(1-4\xi)xz + \frac{2\sqrt{6}}{3}\xi zu$$
$$\left. + 2\xi\beta u\Omega_m - 2\xi\lambda y^2 u + (\gamma-1)\Omega_m \right] \left[ \frac{2}{3} - \frac{2}{3}\xi u^2 - 4\xi^2 u^2 \right]^{-1} . \quad (42)$$

While $s$ is a junk parameter alone, it gains physical meaning in the deceleration parameter $q_D$, such as

$$q_D = -1 + s = -1 + \left[ 1 + (1-4\xi)x^2 - y^2 + (8\xi^2 - \xi)u^2 + \frac{2\sqrt{6}}{3}\xi xu \right.$$
$$+ (1-4\xi)z^2 + 2(1-4\xi)xz$$
$$+ \frac{2\sqrt{6}}{3}\xi zu + 2\xi\beta u\Omega_m - 2\xi\lambda y^2 u$$
$$\left. + (\gamma-1)\Omega_m \right] \left[ \frac{2}{3} - \frac{2}{3}\xi u^2 - 4\xi^2 u^2 \right]^{-1} . \quad (43)$$

Now we convert the Friedmann equations (20), (21), the continuity equation (23), and the equation of motion (33) into the autonomous system by using the auxiliary variables in (34) and their derivatives with respect to $N = \ln a$. For any quantity $F$, this derivative has the relation with the time derivative as $\dot{F} = H(dF/dN) = HF'$. Then we will obtain $X' = f(X)$, where $X$ is the column vector including the auxiliary variables and $f(X)$ is the column vector of the autonomous equations. We then find the critical points $X_c$ of $X$, by setting $X' = 0$. We then expand $X' = f(X)$ around $X = X_c + U$, where $U$ is the column vector of perturbations of the auxiliary variables, such as $\delta x$, $\delta y$, $\delta z$ and $\delta u$ for each constituents in our model. Thus, we expand the perturbation equations up to the first order for each critical point as $U' = MU$, where $M$ is the matrix of perturbation equations. The eigenvalues of perturbation matrix $M$ determine the type and stability of each critical point [101-110]. Then the autonomous form of the cosmological system is

$$x' = -3x - 3z + sx - z' + sz + \sqrt{6}\,\xi\,su - 2\sqrt{6}\,\xi u + \frac{\sqrt{6}}{2}\lambda\,y^2 - \frac{\sqrt{6}}{2}\beta\Omega_m, \tag{44}$$

$$y' = sy - \frac{\sqrt{6}}{2}\lambda\,yx - \frac{\sqrt{6}}{2}\lambda\,yz, \tag{45}$$

$$z' = -3x - 3z + sx - x' + sz + \sqrt{6}\,\xi\,su - 2\sqrt{6}\,\xi u + \frac{\sqrt{6}}{2}\lambda\,y^2 - \frac{\sqrt{6}}{2}\beta\Omega_m, \tag{46}$$

$$u' = \sqrt{6}x + \sqrt{6}z. \tag{47}$$

Here (44) and (46) in fact give the same autonomous equations, which means that the variables $x$ and $z$ do not form an orthonormal basis in the phase-space. However, $x+z$, $y$ and $u$ form a complete orthonormal set for the phase-space. Therefore, we set (44) and (46) in a single autonomous equation as

$$x' + z' = -3x - 3z + sx + sz + \sqrt{6}\,\xi\,su - 2\sqrt{6}\,\xi u + \frac{\sqrt{6}}{2}\lambda\,y^2 - \frac{\sqrt{6}}{2}\beta\Omega_m. \tag{48}$$

The autonomous equation system (45), (47) and (48) represents three invariant submanifolds $x+z=0$, $y=0$ and $u=0$ which, by definition, cannot be intersected by any orbit. This means that there is no global attractor in the deformed dark energy cosmology [111]. We will make finite analysis of the phase space. The finite fixed points are found by setting the derivatives of the invariant submanifolds of the auxiliary variables. We can also write these autonomous equations in $q \to 1$ limit in terms of the standard auxiliary variables, such as

$$x'_s = -3x_s + s_s x_s + \sqrt{6}\,\xi\,s_s u_s - 2\sqrt{6}\,\xi u_s + \frac{\sqrt{6}}{2}\lambda\,y_s^2 - \frac{\sqrt{6}}{2}\beta\Omega_m. \tag{49}$$

$$y'_s = s_s y_s - \frac{\sqrt{6}}{2}\lambda\,y_s x_s, \tag{50}$$

$$u'_s = \sqrt{6}x_s. \tag{51}$$

Here we need to get the finite fixed points (critical points) of the autonomous system in (44)-(48), in order to perform the phase-space analysis of the model. We will obtain these points by equating the left hand sides of the equations (45), (47) and (48) to zero, by using $\Omega_{tot}=1$ in (37) and also by assuming $\omega_{tot}=-1$ and $q_d=-1$ in (40) and (43), for each critical point. After some calculations, four sets of solutions are found as the critical points which are listed in Table 1 with the existence conditions. The same critical points are also valid for the $x_s$, $y_s$ and $u_s$ instead of $x+z$, $y$ and $u$, in the $q \to 1$ standard dark energy model limit.

TABLE 1: Critical points and existence conditions

| Label | $x+z$ | $y$ | $u$ | $\omega_{tot}$ | $q_D$ | Existence |
|---|---|---|---|---|---|---|
| A | 0 | 1 | 0 | $-1$ | $-1$ | $\lambda=0$, $\Omega_m=0$ |
| B | 0 | $-1$ | 0 | $-1$ | $-1$ | $\lambda=0$, $\Omega_m=0$ |
| C | 0 | $\sqrt{\dfrac{4\xi}{\lambda}\left(-\dfrac{2}{\lambda}+\sqrt{\dfrac{4}{\lambda^2}+\dfrac{1}{\xi}}\right)}$ | $\left(-\dfrac{2}{\lambda}+\sqrt{\dfrac{4}{\lambda^2}+\dfrac{1}{\xi}}\right)$ | $-1$ | $-1$ | $\lambda \neq 0$, $\Omega_m=0$ |
| D | 0 | $-\sqrt{\dfrac{4\xi}{\lambda}\left(-\dfrac{2}{\lambda}+\sqrt{\dfrac{4}{\lambda^2}+\dfrac{1}{\xi}}\right)}$ | $\left(-\dfrac{2}{\lambda}+\sqrt{\dfrac{4}{\lambda^2}+\dfrac{1}{\xi}}\right)$ | $-1$ | $-1$ | $\lambda \neq 0$, $\Omega_m=0$ |

Now we should find $\delta s$ from (42), which will exist in the perturbations $\delta x' + \delta z'$, $\delta y'$ and $\delta u'$, such that

$$\delta s = \left[2(1-4\xi)(x+z)+\frac{2\sqrt{6}}{3}\xi u\right]\frac{1}{P}(\delta x+\delta z)+\left[-2y-4\xi\lambda yu\right]\frac{1}{P}\delta y$$
$$+\left[(8\xi^2-\xi)2u+\frac{2\sqrt{6}}{3}\xi(x+z)+2\xi\beta\Omega_m-2\xi\lambda y^2+(\frac{4}{3}\xi+8\xi^2)su\right]\frac{1}{P}\delta u, \quad (52)$$

where $P = 2/3 - (2/3)\xi u^2 - 4\xi^2 u^2$. Then the perturbations $\delta x' + \delta z'$, $\delta y'$ and $\delta u'$ for each phase-space coordinates in our model can be found by using the variations of the equations (45), (47) and (48), such as

$$\delta x' + \delta z' = \left[ 2(1-4\xi)(x+z)^2 + (-s+3)P + 4\xi^2 u^2 + (\frac{8}{3}\xi - 8\xi^2)\sqrt{6}(x+z)u \right] \frac{1}{P}(\delta x + \delta z)$$

$$+ \left[ (-2(x+z) + \sqrt{6}\lambda P)y - 4\xi\lambda(x+z)yu - 2\sqrt{6}\xi yu - 4\sqrt{6}\xi^2 \lambda yu^2 \right] \frac{1}{P} \delta y$$

$$+ \left[ (10\xi^2 - \xi)2(x+z)u + \frac{2\sqrt{6}}{3}\xi(x+z)^2 + 2\xi\beta(x+z)\Omega_m \right.$$

$$- 2\xi\lambda(x+z)y^2 + (\frac{4}{3}\xi + 8\xi^2)(x+z)su + (8\xi^2 - \xi)2\sqrt{6}\xi u^2$$

$$+ 2\sqrt{6}\xi^2 \beta u\Omega_m - 2\sqrt{6}\xi^2 \lambda y^2 u + (\frac{4}{3}\xi + 8\xi^2)\sqrt{6}\xi su^2 + \sqrt{6}\xi(s-2)P \left.\right] \frac{1}{P} \delta u$$

(53)

$$\delta y' = \left[ 2(1-4\xi)(x+z)y + \frac{2\sqrt{6}}{3}\xi uy - \frac{\sqrt{6}}{2}\lambda yP \right] \frac{1}{P}(\delta x + \delta z)$$

$$+ \left[ -2y^2 - 4\xi\lambda y^2 u - +sP - \frac{\sqrt{6}}{2}\lambda(x+z)P \right] \frac{1}{P} \delta y$$

$$+ \left[ (8\xi^2 - \xi)uy + \frac{2\sqrt{6}}{3}\xi(x+z)y + 2\xi\beta y\Omega_m \right.$$

$$- 2\xi\lambda y^3 + (\frac{4}{3}\xi + 8\xi^2)syu \left.\right] \frac{1}{P} \delta u$$

(54)

$$\delta u' = \sqrt{6}(\delta x + \delta z).$$

(55)

From (53)-(55), we find the $3\times 3$ perturbation matrix $M$ whose elements are given as

$$M_{11} = \left[ 2(1-4\xi)(x+z)^2 + (-s+3)P + 4\xi^2 u^2 + (\frac{8}{3}\xi - 8\xi^2)\sqrt{6}(x+z)u \right] \frac{1}{P},$$

$$M_{12} = \left[ (-2(x+z) + \sqrt{6}\lambda P)y - 4\xi\lambda(x+z)yu - 2\sqrt{6}\xi yu - 4\sqrt{6}\xi^2 \lambda yu^2 \right] \frac{1}{P},$$

$$M_{13} = \left[ (10\xi^2 - \xi)2(x+z)u + \frac{2\sqrt{6}}{3}\xi(x+z)^2 + 2\xi\beta(x+z)\Omega_m \right.$$

$$- 2\xi\lambda(x+z)y^2 + (\frac{4}{3}\xi + 8\xi^2)(x+z)su + (8\xi^2 - \xi)2\sqrt{6}\,\xi u^2$$

$$+ 2\sqrt{6}\,\xi^2 \beta u\Omega_m - 2\sqrt{6}\,\xi^2 \lambda\, y^2 u + (\frac{4}{3}\xi + 8\xi^2)\sqrt{6}\xi\, su^2$$

$$\left. + \sqrt{6}\,\xi(s-2)P \right]\frac{1}{P}$$

$$M_{21} = \left[ 2(1-4\xi)(x+z)y + \frac{2\sqrt{6}}{3}\xi uy - \frac{\sqrt{6}}{2}\lambda yP \right]\frac{1}{P},$$

$$M_{22} = \left[ -2y^2 - 4\xi\lambda\, y^2 u - +sP - \frac{\sqrt{6}}{2}\lambda(x+z)P \right]\frac{1}{P},$$

$$M_{23} = \left[ (8\xi^2 - \xi)uy + \frac{2\sqrt{6}}{3}\xi(x+z)y + 2\xi\beta\, y\Omega_m \right.$$

$$\left. - 2\xi\lambda\, y^3 + (\frac{4}{3}\xi + 8\xi^2)syu \right]\frac{1}{P}$$

$$M_{31} = \sqrt{6},$$

$$M_{32} = M_{33} = 0. \tag{56}$$

We insert the linear perturbations $(x+z) \to (x_c + z_c) + (\delta x + \delta z)$, $y \to y_c + \delta y$ and $u \to u_c + \delta u$ about the critical points in the autonomous system (45), (47) and (48), in order to calculate the eigenvalues of perturbation matrix $M$ for four critical points given in Table 1, with the corresponding existing conditions. Therefore, we first give the four perturbation matrices for the critical points $A$, $B$, $C$ and $D$ with the corresponding existing conditions, such as

$$M_A = M_B = \begin{pmatrix} -3 & 0 & -2\sqrt{6}\xi \\ 0 & -3 & 0 \\ \sqrt{6} & 0 & 0 \end{pmatrix}, \tag{57}$$

$$M_C = \begin{pmatrix} \frac{4\xi^2 u_C^2}{P} - 3 & -\frac{4\sqrt{6}\lambda u_C^2 \xi^2}{P} - \frac{2\sqrt{6}\xi y_C u_C}{P} + \sqrt{6} y_C \lambda & -\frac{\sqrt{6}(2\xi^2 - 16\xi^3)u_C^2}{P} - \frac{2\sqrt{6}\xi^2 \lambda y_C^2 u_C}{P} - 2\sqrt{6}\xi \\ \frac{2\sqrt{6}\xi y_C u_C}{3P} - \frac{\sqrt{6}\lambda y_C}{2} & -\frac{2 y_C^2}{P} - \frac{4\xi \lambda y_C^2 u_C}{P} & -\frac{2\xi \lambda y_C^3}{P} - \frac{(2\xi - 16\xi^2)u_C y_C}{P} \\ \sqrt{6} & 0 & 0 \end{pmatrix},$$

(58)

where $y_C = \sqrt{4\xi/\lambda(-2/\lambda + \sqrt{4/\lambda^2 + 1/\xi})}$ and $u_C = -2/\lambda + \sqrt{4/\lambda^2 + 1/\xi}$.

$$M_D = \begin{pmatrix} \frac{4\xi^2 u_D^2}{P} - 3 & \frac{4\sqrt{6}\lambda u_D^2 \xi^2}{P} + \frac{2\sqrt{6}\xi y_D u_D}{P} - \sqrt{6} y_D \lambda & -\frac{\sqrt{6}(2\xi^2 - 16\xi^3)u_D^2}{P} - \frac{2\sqrt{6}\xi^2 \lambda y_D^2 u_D}{P} - 2\sqrt{6}\xi \\ \frac{\sqrt{6}\lambda y_D}{2} - \frac{2\sqrt{6}\xi y_D u_D}{3P} & -\frac{2 y_D^2}{P} - \frac{4\xi \lambda y_D^2 u_D}{P} & -\frac{2\xi \lambda y_D^3}{P} - \frac{(2\xi - 16\xi^2)u_D y_D}{P} \\ \sqrt{6} & 0 & 0 \end{pmatrix},$$

(59)

where $y_D = -\sqrt{4\xi/\lambda(-2/\lambda + \sqrt{4/\lambda^2 + 1/\xi})}$ and $u_D = -2/\lambda + \sqrt{4/\lambda^2 + 1/\xi}$. Also by using $x_s$, $y_s$ and $u_s$ instead of $x+z$, $y$ and $u$ in the perturbation matrix elements above, we obtain the standard perturbation matrix elements in $q \to 1$ limit. Then substituting the standard critical points we again obtain the same matrices $M_A$, $M_B$, $M_C$ and $M_D$. Therefore the stability of the standard model agrees with the stability of the deformed model.

We need to obtain the four sets of eigenvalues and investigate the sign of the real parts of eigenvalues, so that we can determine the type and stability of critical points. If the all real parts of the eigenvalues are negative, the critical point is said to be stable. The physical meaning of the stable critical point is a stable attractor, namely the universe keeps its state forever in this state and thus it can attract the universe at a late-time. Here an accelerated expansion phase occurs, because $\omega_{tot} = -1 < -1/3$. However, if the suitable conditions are satisfied, there can even exist an accelerated contraction for $\omega_{tot} = -1 < -1/3$ value. Eigenvalues of the four $M$ matrices and the stability conditions are represented in Table 2, for each critical points $A$, $B$, $C$ and $D$. From Table 2, the first two critical points $A$ and $B$ have the same eigenvalues, as $C$ and $D$ have the same eigenvalues, too. Here the eigenvalues and the stability conditions of the perturbation matrices for critical points $C$ and $D$ have been obtained by the numerical methods, since the complexity of the matrices (58) and (59). The stability conditions of each critical point are listed in Table 2, according to the sign of the real part of the eigenvalues.

TABLE 2: Eigenvalues and stability of critical points

| Critical Points | Eigenvalues | Stability |
|---|---|---|
| A and B | -3.0000,  $-\frac{1}{2}\sqrt{9-48\xi}-\frac{3}{2}$,  $\frac{1}{2}\sqrt{9-48\xi}-\frac{3}{2}$ | Stable point for $0<\xi\leq\frac{3}{16}$ with $\lambda,\beta\in\Re$<br>Saddle point for $\xi<0$ with $\lambda,\beta\in\Re$ |
| C and D | Eigenvalues  $\xi$  $\lambda$<br>-1.0642  -1.5576  -5.5000   0.1000  1.0000<br>-1.0193  -1.0193  -7.2507   1.0000  1.0000<br>-0.8407  -0.8407  -7.7519   2.0000  1.0000<br>-0.7080  -0.7080  -8.0701   3.0000  1.0000<br>-0.6014  -0.6014  -8.3107   4.0000  1.0000<br>-0.5121  -0.5121  -8.5060   5.0000  1.0000<br>-0.4353  -0.4353  -8.6709   6.0000  1.0000<br>-0.3680  -0.3680  -8.8136   7.0000  1.0000<br>-0.3082  -0.3082  -8.9395   8.0000  1.0000<br>-0.2544  -0.2544  -9.0520   9.0000  1.0000<br>-0.2055  -0.2055  -9.1535  10.0000  1.0000 | Stable point for $0<\xi$, $\lambda=1$ and $\beta\in\Re$<br>Saddle point, if $\xi<0$ and $\lambda\neq1$ |

Now we will study the cosmological behavior of each critical point by considering the attractor solutions in scalar field cosmology [112]. We know that the energy density of a scalar field has a role on the determination of the evolution of universe. Cosmological attractors provide the understanding of evolution and the affecting factors on this evolution, such as, from the dynamical conditions that the scalar field evolution approaches a certain kind of behavior without initial fine tuning conditions [113–123]. We know that the attractor solutions imply a behavior in which a collection of phase-space points evolve into a particular region and never leave from there. In order to solve the differential equation system (45), (47) and (48) we use adaptive Runge-Kutta method of 4th and 5th order, in Matlab programming. We use the present day values for the dark matter density parameter $\Omega_m=0.3$, interaction parameter $\beta=14.5$ and $0<\gamma<2$ values in solving the differential equation system [96, 124]. Then the solutions with the stability conditions of critical points are plotted for each pair of the solution set being the auxiliary variables $x+z$, $y$ and $u$.

*Critical point A:* This point exists for $\lambda = 0$ which means the potential $V$ is constant. Acceleration occurs at this point because of $\omega_{tot} = -1 < -1/3$, and it is an expansion phase since $y$ is positive, so $H$ is positive, too. Point $A$ is stable meaning that universe keeps its further evolution, for $0 < \xi \leq 3/16$ with $\lambda, \beta \in \Re$, but it is a saddle point meaning the universe evolves between different states for $\xi < 0$. In Figure 1, we illustrate the 2-dimensional projections of 4-dimensional phase-space trajectories for the stability condition $\xi = 0.15$ and for the present day values $\beta = 14.5$, $\gamma = 1.5$, $\Omega_m = 0.3$ and three auxiliary $\lambda$ values. This state corresponds to a stable attractor starting from the critical point $A = (0,1,0)$, as seen from the plots in Figure 1.

*Critical point B:* Point $B$ also exists for $\lambda = 0$ meaning the potential $V$ is constant. Acceleration phase is again valid here since $\omega_{tot} = -1 < -1/3$, but this point refers to contraction phase because $y$ is negative here. For the stability of the point $B$, it is again stable for $0 < \xi \leq 3/16$ with $\lambda, \beta \in \Re$, but it is a saddle point for $\xi < 0$. Therefore it is represented that the stable attractor behavior for contraction starting from the critical point $B = (0,-1,0)$, as seen from the graphs in Figure 2. We plot phase-space trajectories for the stability condition $\xi = 0.15$ and for the present day values $\beta = 14.5$, $\gamma = 1.5$, $\Omega_m = 0.3$ and three auxiliary $\lambda$ values.

*Critical point C:* Critical point $C$ occurs for $\lambda \neq 0$ meaning a field dependent potential $V$. The cosmological behavior is again an acceleration phase since $\omega_{tot} < -1/3$ and an expansion phase since $y$ is positive. Point $C$ is stable for $0 < \xi$, $\lambda = 1$ and $\beta \in \Re$, and saddle point if $\xi < 0$ and $\lambda \neq 1$. 2-dimensional projections of phase-space are represented in Figure 3, for the stability conditions $\xi = 2$, $\lambda = 1$ and for the present day values $\beta = 14.5$, $\Omega_m = 0.3$ and three auxiliary $\gamma$ values in the present day value range. The stable attractor starting from the critical point $C$ can be inferred from the plots in Figure 3.

*Critical point D:* This point exists for $\lambda \neq 0$ meaning a field dependent potential $V$. Acceleration phase is again valid due to $\omega_{tot} < -1/3$, but this point refers to a contraction phase, because $y$ is negative. Point $D$ is also stable for $0 < \xi$, $\lambda = 1$ and $\beta \in \Re$. However, it is a saddle point, while $\xi < 0$ and $\lambda \neq 1$. 2-dimensional plots of phase-space trajectories are shown in Figure 4, for the stability conditions $\xi = 2$, $\lambda = 1$ and for the present day values $\beta = 14.5$, $\Omega_m = 0.3$ and three auxiliary $\gamma$ values in the present day value range. This state again corresponds to a stable attractor starting from the point $D$, as seen from the plots in Figure 4.

FIGURE 1: Two dimensional projections of the phase-space trajectories for stability condition $\xi = 0.15$ and for present day values $\beta = 14.5$, $\gamma = 1.5$, $\Omega_m = 0.3$. All plots begin from the critical point $A$ being a stable attractor.

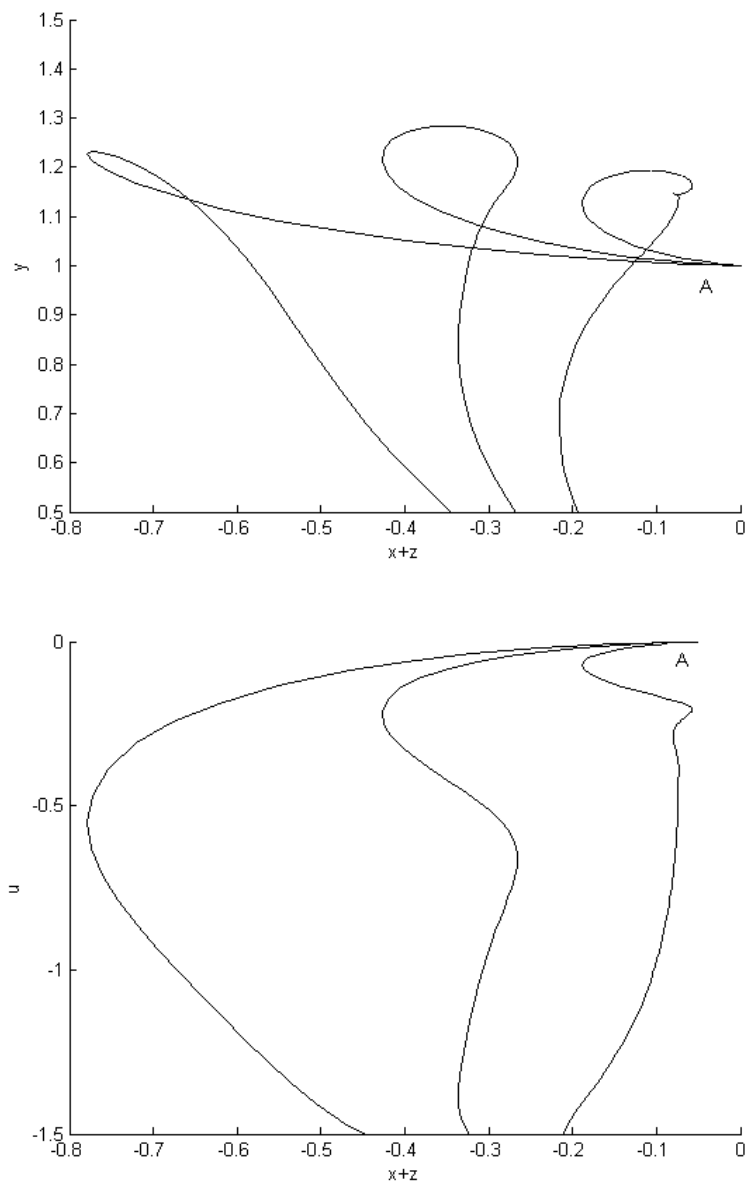

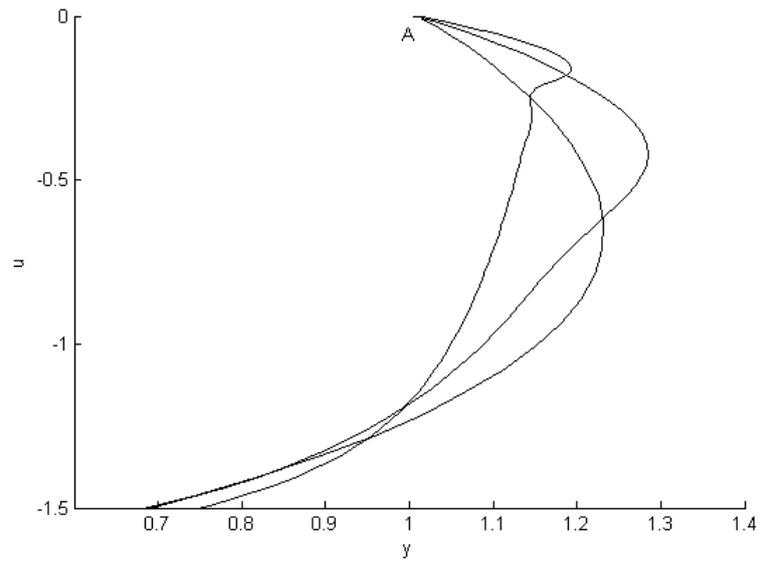

FIGURE 2: Two dimensional projections of the phase-space trajectories for stability condition $\xi = 0.15$ and for present day values $\beta = 14.5$, $\gamma = 1.5$, $\Omega_m = 0.3$. All plots begin from the critical point $B$ being a stable attractor.

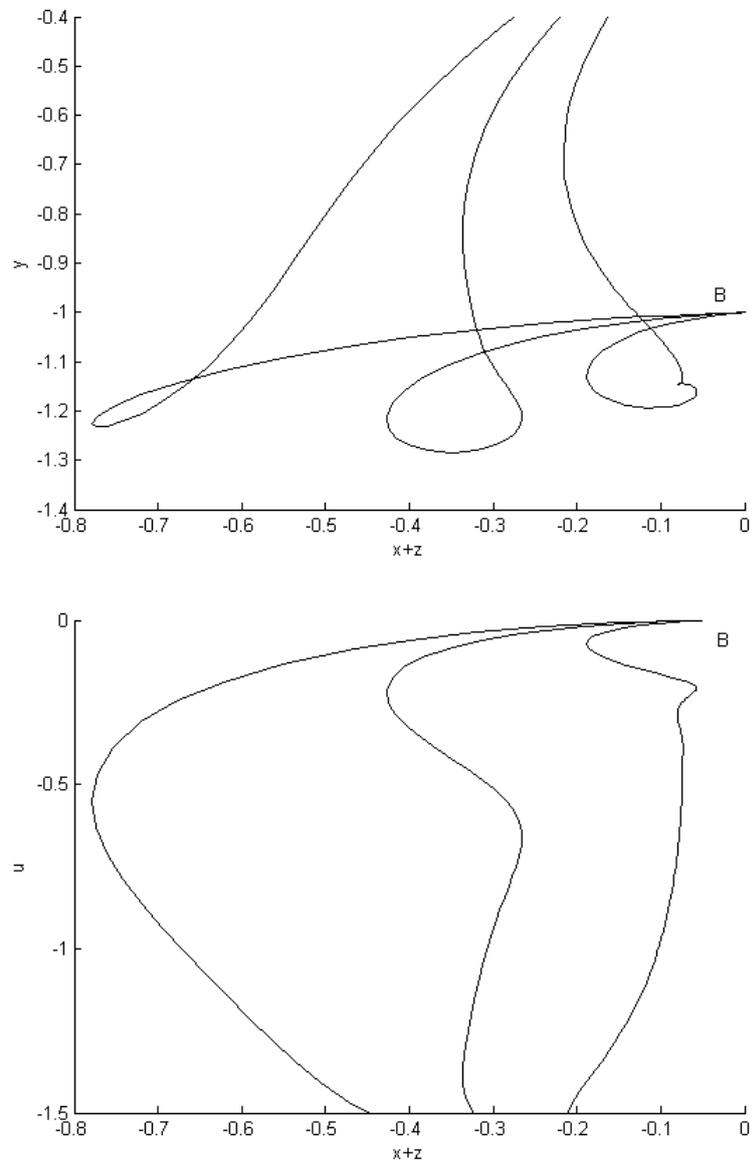

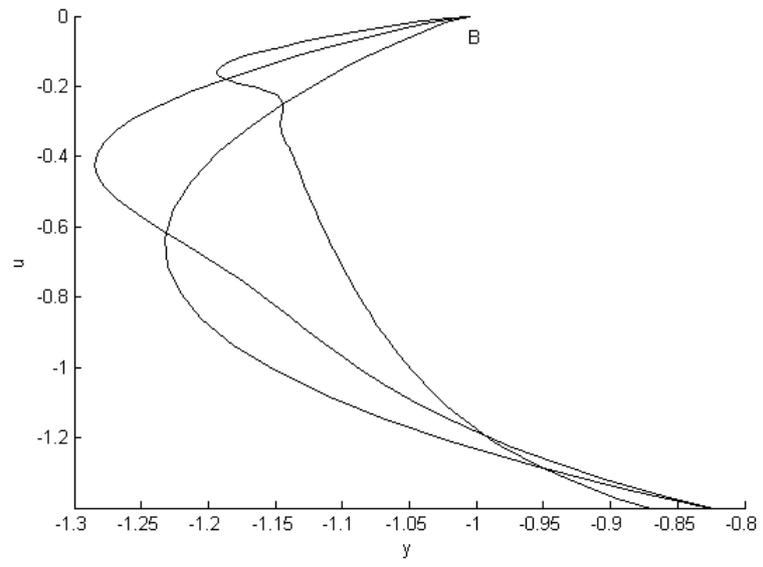

FIGURE 3: Two dimensional projections of the phase-space trajectories for stability conditions $\xi = 2$, $\lambda = 1$ and for present day values $\beta = 14.5$, $\Omega_m = 0.3$. All plots begin from the critical point $C$ being a stable attractor.

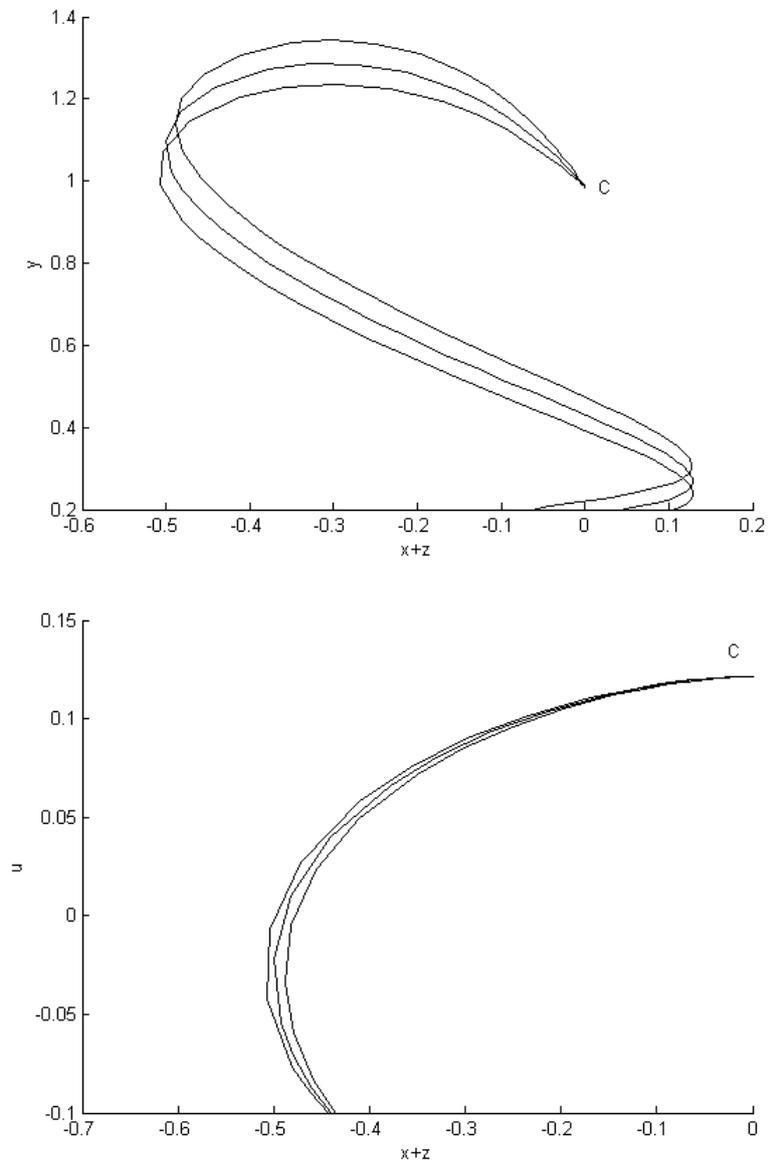

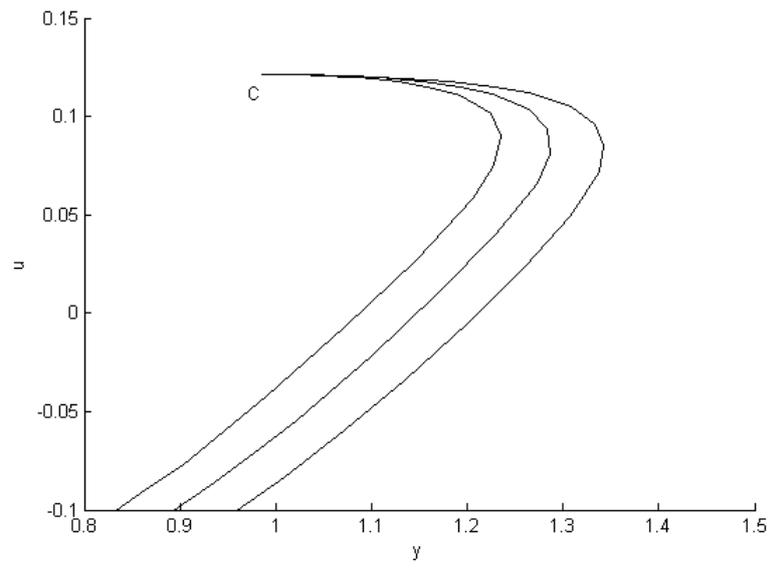

FIGURE 4: Two dimensional projections of the phase-space trajectories for stability conditions $\xi = 2$, $\lambda = 1$ and for present day values $\beta = 14.5$, $\Omega_m = 0.3$. All plots begin from the critical point $D$ being a stable attractor.

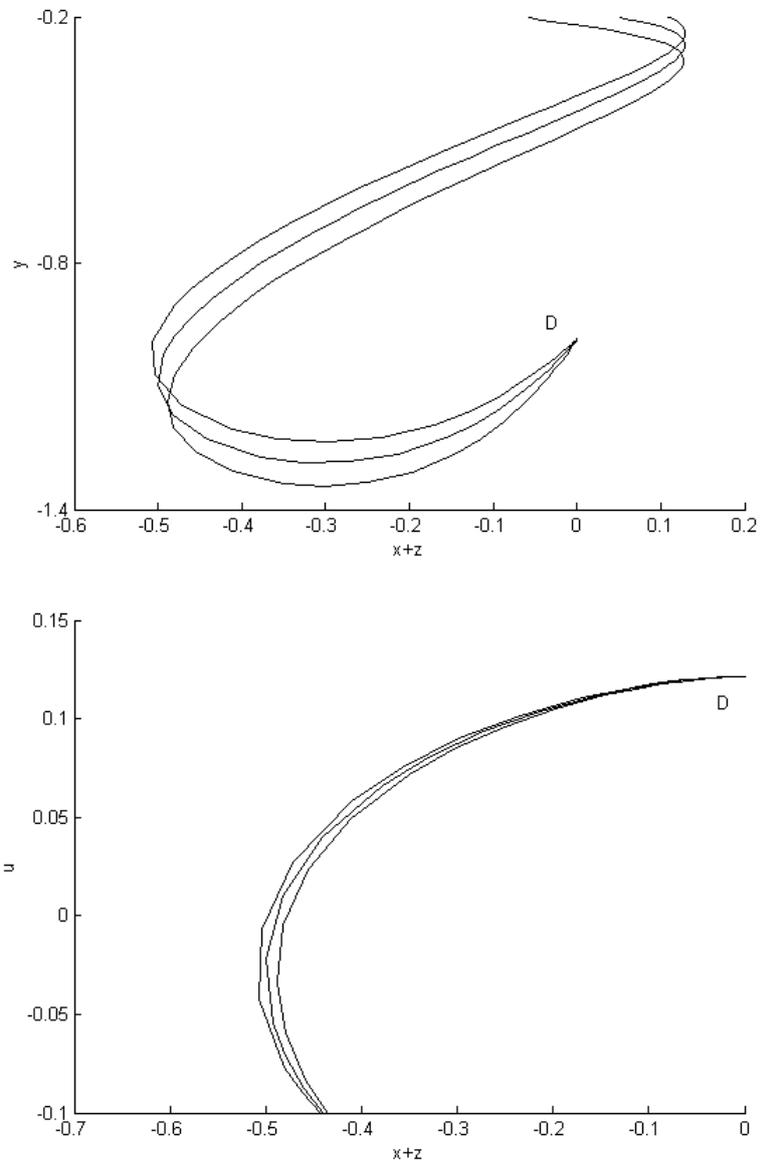

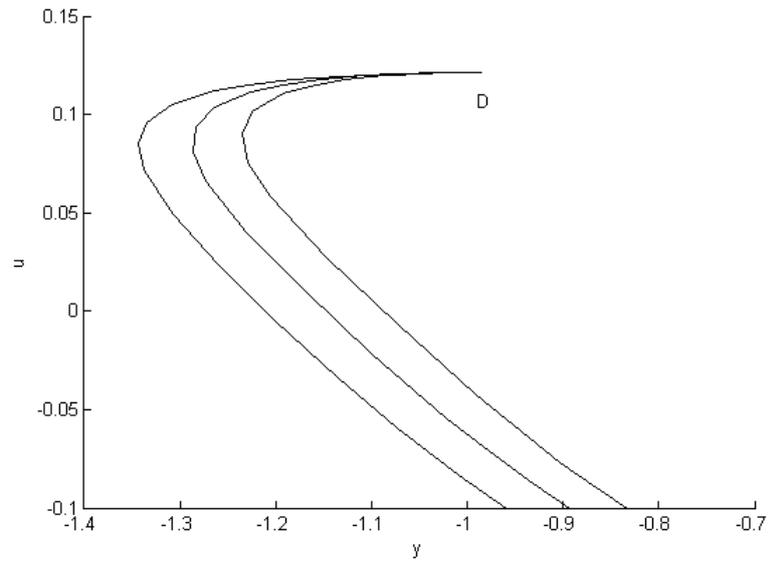

FIGURE 5: Three dimensional plots of the phase-space trajectories for the critical point $A$, $B$, $C$ and $D$ being the stable attractors.

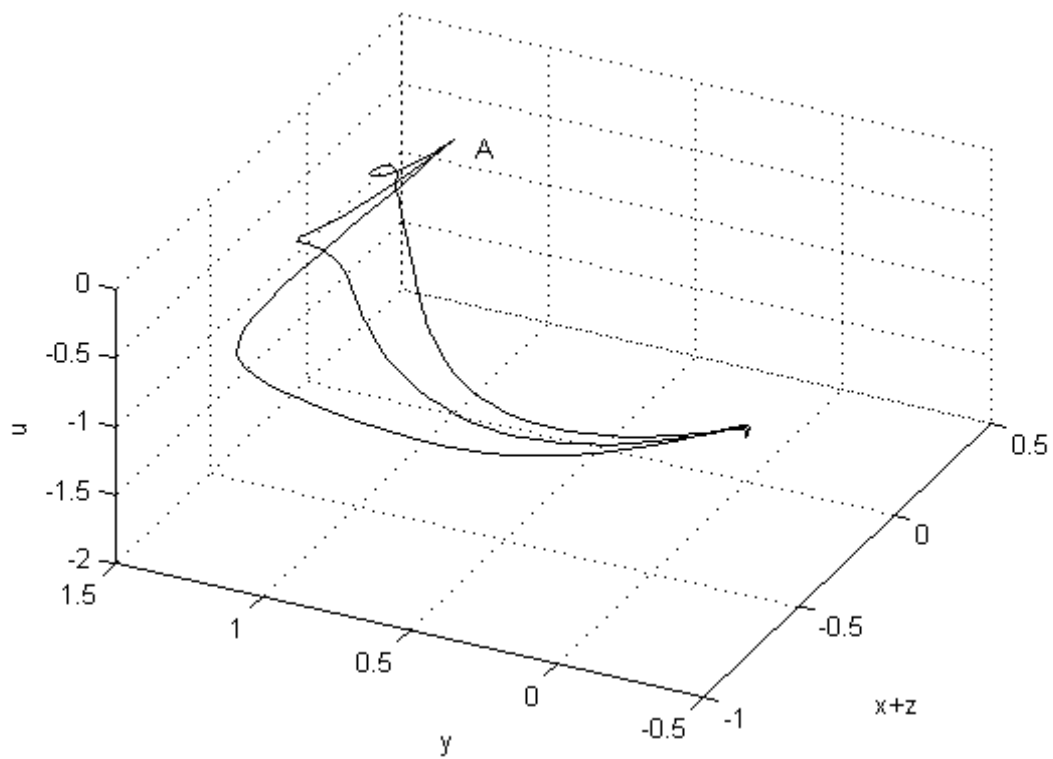

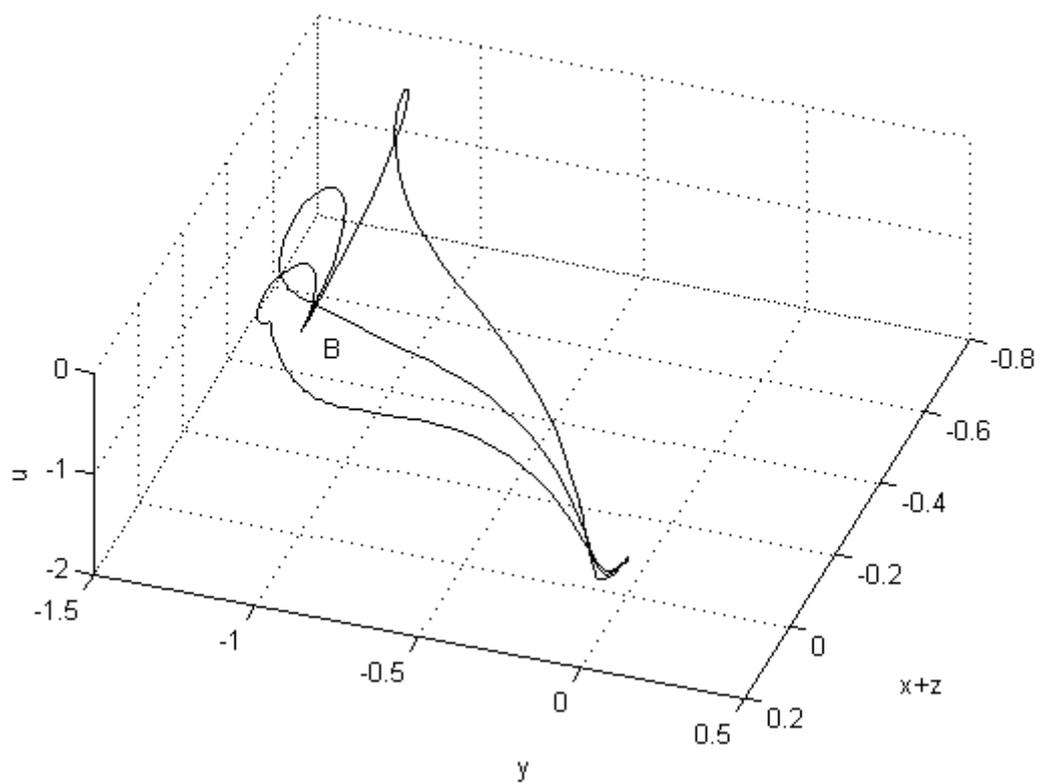

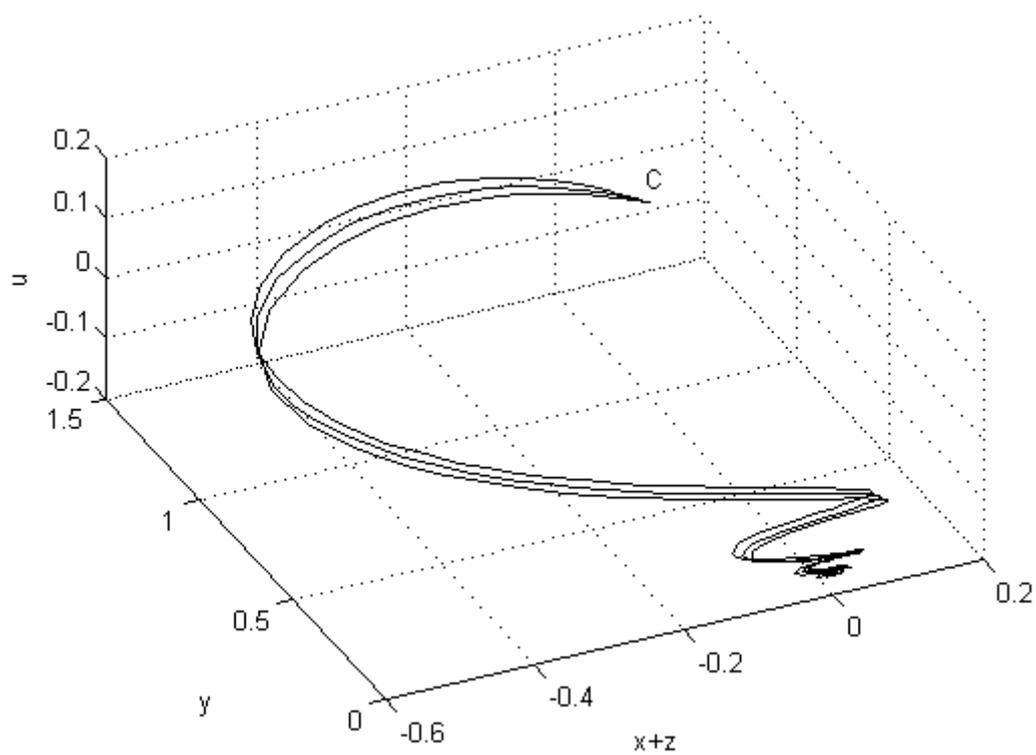

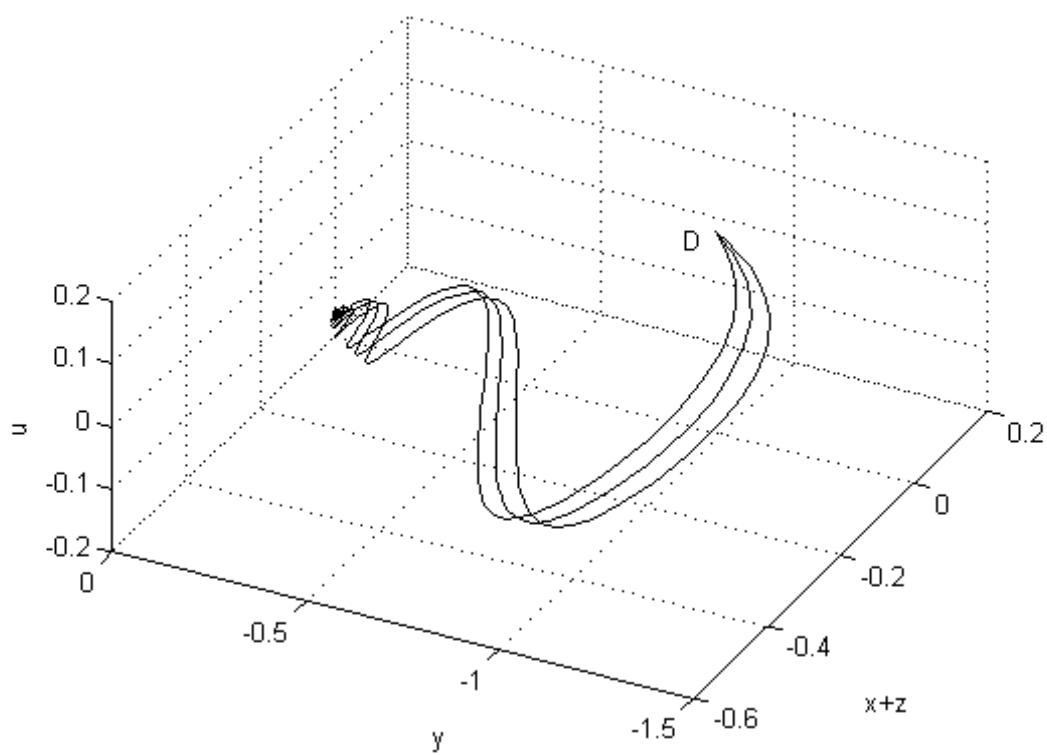

All the plots in Figures 1-4 have the structure of stable attractor, since each of them evolves to a single point which is in fact one of the critical points in Table 1. The three dimensional plots of the evolution of phase-space trajectories for the stable attractors are given in Figures 5. These evolutions to the critical points are the attractor solutions of our cosmological model: interacting dark matter and $q$-deformed dark energy non-minimally coupled to gravity, which imply an expanding universe. On the other hand, the construction of the model in the $q \to 1$ limit reproduces the results of the phase space analysis for the non-deformed standard dark energy case. The critical points, perturbation matrices are same for the deformed and standard dark energy models with the equivalence of the auxiliary variables as $x + z = x_s$, $y = y_s$ and $u = u_s$. Therefore, it is confirmed that the dark energy in our model can be defined in terms of the $q$-deformed scalar fields obeying the $q$-deformed boson algebra in (2) and (3). According to the stable attractor behaviors, it makes sense to consider the dark energy as a scalar field defined by the $q$-deformed scalar field, since the negative pressure of $q$-deformed boson field, as dark energy field.

We know that the deformed dark energy model is a confirmed model since it reproduces the same stability behaviors, critical points and perturbation matrices with the standard dark energy model, but the auxiliary variables of deformed and standard models are not same. The relation between deformed and standard dark energy can be represented regarding to auxiliary variable equations in (34).

$$x = \sqrt{\frac{1-q^{2N}}{(1-q^2)N}} x_s, \quad y = \sqrt{\exp\left(-c\left(\sqrt{\frac{1-q^{2N}}{(1-q^2)N}} - 1\right)\right)} y_s, \quad u = \sqrt{\frac{1-q^{2N}}{(1-q^2)N}} u_s, \quad (60)$$

where $c$ is a constant. From the equations (60) we now illustrate the behavior of the deformed and standard dark energy auxiliary variables with respect to the deformation parameter $q$ in Figure 6. We infer from the figure that the value of the deformed $x$, $y$ and $u$ decreases with decreasing $q$ for the $q < 1$ interval for large particle number, and the decrease in the variables $x$, $y$ and $u$ refer to the decrease in deformed energy density. Also, we conclude that the value of the auxiliary variables $x$, $y$ and $u$ increases with increasing $q$ for the $q > 1$ interval for large particle number. In $q \to 1$ limit deformed variables goes to standard ones.

FIGURE 6: Behavior of the auxiliary variables $x$, $y$ and $u$ with respect to the deformation parameter $q$ and the particle number $N$.

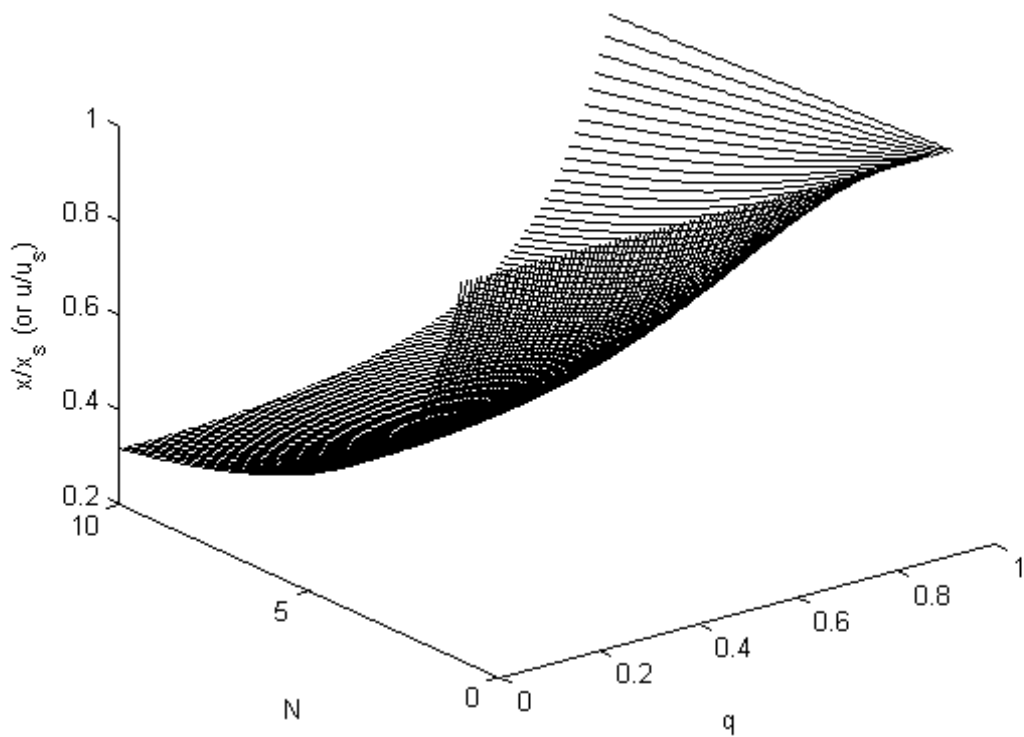

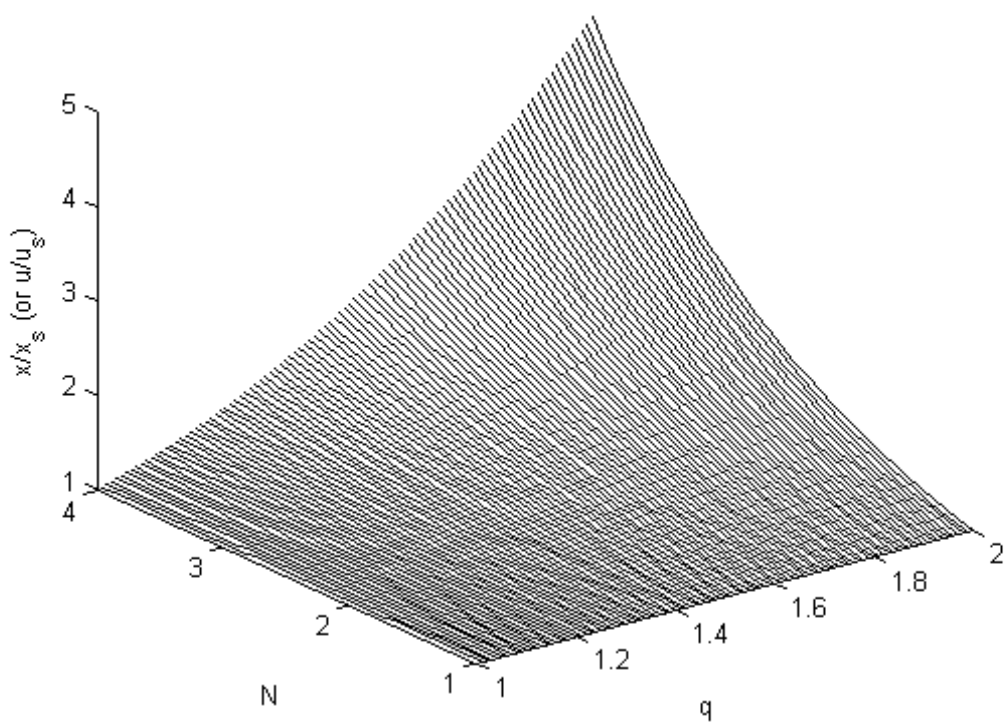

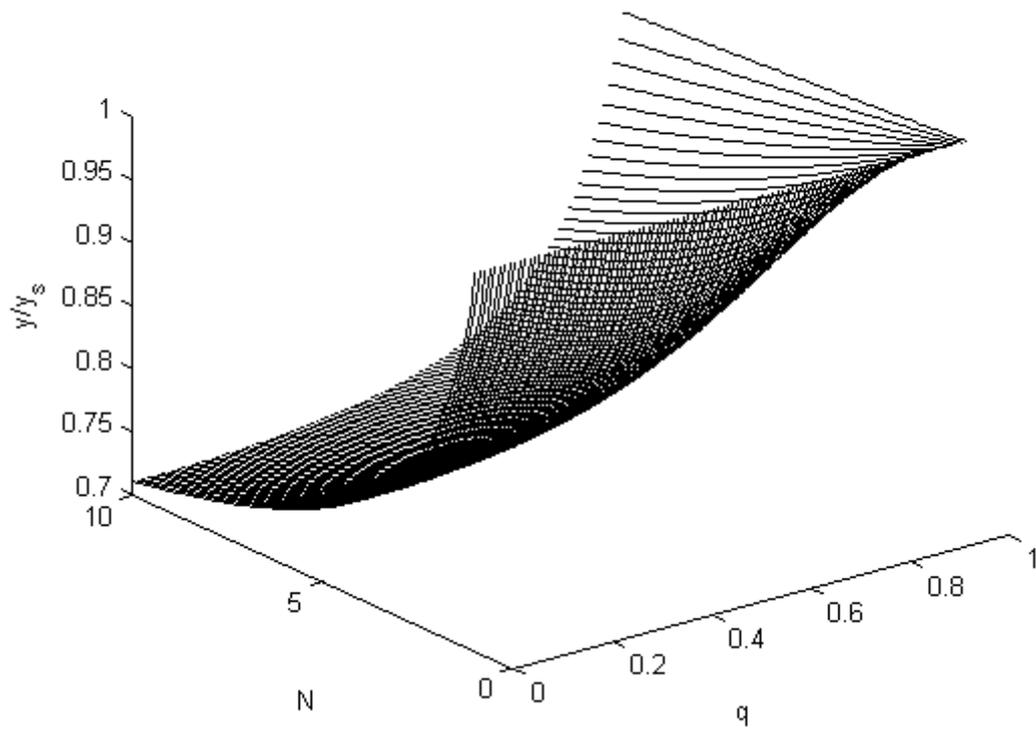

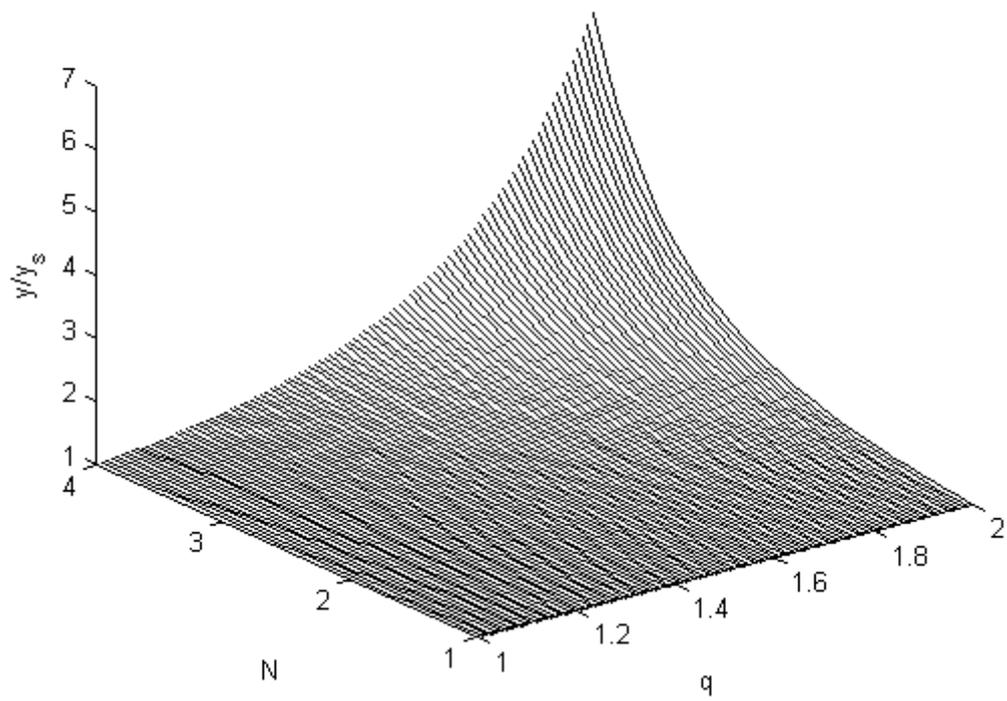

## 4. Conclusion

In this study, we propose that the dark energy is formed of the negative-pressure $q$-deformed scalar field whose field equation is defined by the $q$-deformed annihilation and creation operators satisfying the deformed boson algebra in (2) and (3), since it is known that the dark energy has a negative pressure - like the deformed bosons - acting as a gravitational repulsion to drive the accelerated expansion of universe. We consider an interacting dark matter and $q$-deformed dark energy non-minimally coupled to the gravity in the framework of Einsteinian gravity in order to confirm our proposal. Then we investigate the dynamics of the model and phase-space analysis whether it will give stable attractor solutions meaning indirectly an accelerating expansion phase of universe. Therefore, we construct the action integral of the interacting dark matter and $q$-deformed dark energy non-minimally coupled to gravity model in order to study its dynamics. With this the Hubble parameter and Friedmann equations of the model are obtained in the spatially-flat FRW geometry. Later on, we find the energy density and pressure with the evolution equations for the $q$-deformed dark energy and dark matter from the variation of the action and the Lagrangian of the model. After that we translate these dynamical equations into the autonomous form by introducing the suitable auxiliary variables, in order to perform the phase-space analysis of the model. Then the critical points of autonomous system are obtained by setting each autonomous equation to zero and four perturbation matrices are obtained for each critical point by constructing the perturbation equations. We then determine the eigenvalues of four perturbation matrices to examine the stability of the critical points. We also calculate some important cosmological parameters, such as the total equation of state parameter and the deceleration parameter to check whether the critical points satisfy an accelerating universe. We obtain four stable attractors for the model depending on the coupling parameter $\xi$, interaction parameter $\beta$ and the potential constant $\lambda$. An accelerating universe exists for all stable solutions due to $\omega_{tot} < -1/3$. The critical points $A$ and $B$ are late-time stable attractors for $0 < \xi \leq 3/16$ and $\lambda, \beta \in \Re$, with the point $A$ refers to an expansion with a stable acceleration, while the point $B$ refers to a contraction. However, the critical points $C$ and $D$ are late-time stable attractors for $0 < \xi$, $\lambda = 1$ and $\beta \in \Re$, with the point $C$ refers to an expansion with a stable acceleration, while the point $D$ refers to a contraction. The stable attractor behavior of the model at each critical point is demonstrated in Figures 1-4. In order to solve the differential equation system (45), (47) and (48) with the critical points and plot the graphs in Figures 1-4, we use adaptive Runge-Kutta method of 4th and 5th order, in Matlab programming. Then the solutions with the stability conditions of critical points are plotted for each pair of the solution set being the auxiliary variables in $x+z$, $y$ and $u$.

These figures show that by using the convenient parameters of the model according to the existence and stability conditions and the present day values, we can obtain the stable attractors as $A$, $B$, $C$ and $D$.

The $q$-deformed dark energy is a generalization of the standard scalar field dark energy. As seen from (10) in the $q \to 1$ limit, the behavior of the deformed energy density, pressure and scalar field functions with respect to the standard functions, they all approach to the standard corresponding function values. Consequently, $q$ deformation of the scalar field dark energy gives a self-consistent model due to the existence of standard case parameters of the dark energy in the $q \to 1$ limit and the existence of the stable attractor behavior of the accelerated expansion phase of universe for the considered interacting and non-minimally coupled dark energy and dark matter model. Although the deformed dark energy model is confirmed through reproducing the same stability behaviors, critical points and perturbation matrices with the standard dark energy model, the auxiliary variables of deformed and standard models are of course different. By using the auxiliary variable equations in (34), we find the relation between deformed and standard dark energy variables. From these equations, we represent the behavior of the deformed and standard dark energy auxiliary variables with respect to the deformation parameter for $q<1$ and $q>1$ intervals in Figure 6. Then, the value of the deformed $x$, $y$ and $u$ or equivalently deformed energy density decreases with decreasing $q$ for the $q<1$ interval for large particle number. Also, value of the auxiliary variables $x$, $y$ and $u$ increases with increasing $q$ for the $q>1$ interval for large particle number. In $q \to 1$ limit all the deformed variables goes to non-deformed variables.

The consistency of the proposed $q$-deformed scalar field dark energy model is confirmed by the results, since it gives the expected behavior of the universe. The idea to consider the dark energy as a $q$-deformed scalar field is a very recent approach. There are more deformed particle algebras in the literature which can be considered as other and maybe more suitable candidates for the dark energy. As a further study for the confirmation whether the dark energy can be considered as a general deformed scalar field, the other interactions and couplings between deformed dark energy models, dark matter and gravity can be investigated in the general relativity framework or in the framework of other modified gravity theories, such as teleparallel.

**Conflict of Interest**

The author declares that there is no conflict of interest regarding the publication of this paper.